\documentclass[preprint,aps,showpacs,amsmath,amssymb,superscriptaddress]{revtex4}

\usepackage{graphicx}
\usepackage{dcolumn}
\usepackage{bm}
\usepackage{subfigure}

\begin{document}


\title{Mott transitions of exciton-polaritons and indirect excitons in a periodic potential}
\author{Tim Byrnes}
\affiliation{National Institute of Informatics, 2-1-2
Hitotsubashi, Chiyoda-ku, Tokyo 101-8430, Japan}
\affiliation{Institute of Industrial Science,
University of Tokyo, 4-6-1 Komaba, Meguro-ku, Tokyo 153-8505,
Japan} 

\author{Patrik Recher}
\affiliation{Institut f\"{u}r Theoretische Physik und Astrophysik, Universit\"{a}t W\"{u}rzburg, \\Am Hubland, D--97074 W\"{u}rzburg, Germany}

\author{Yoshihisa Yamamoto}
\affiliation{National Institute of Informatics, 2-1-2
Hitotsubashi, Chiyoda-ku, Tokyo 101-8430, Japan}
\affiliation{E.
L. Ginzton Laboratory, Stanford University, Stanford, CA 94305}
\date{\today}

\date{\today}

\begin{abstract}
We derive an effective Bose-Hubbard model that predicts a phase transition from Bose-Einstein 
condensate to Mott insulator in two different systems subject to applied periodic potentials: 
microcavity exciton-polaritons and indirect excitons.
Starting from a microscopic Hamiltonian of electrons and holes, we derive 
an effective Bose-Hubbard model for both systems and evaluate the on-site Coulomb interaction $ U $ and 
hopping transition amplitudes $ t $. Experimental parameters required for observing a phase transition
between a Bose-Einstein condensate and a Mott insulator are discussed. Our results suggest that strong 
periodic potentials and polaritons with a very large excitonic component are required for observing the phase transition. 
The form of the indirect exciton interaction is derived including direct and exchange components of the Coulomb interaction. 
For indirect excitons, the system crosses over from a Bose-Hubbard model into a double layer Fermi-Hubbard 
model as a function of increasing bilayer separation. The Fermi-Hubbard model parameters are calculated, and 
the criteria for the location of this crossover are derived. We conjecture that a crossover between 
a Bose Mott insulator to a Fermi Mott insulator should occur with increasing bilayer separation. 
\end{abstract}

\maketitle

\section{Introduction}

The observation of the Bose-Einstein condensation (BEC) of exciton-polaritons has generated a large 
amount of interest in recent years \cite{deng02,kasprzak06,balili07}. The focus has now turned to 
examining various properties of the condensate, such as thermal equilibration \cite{deng06},
superfluidity \cite{amo09}, vortex formation \cite{lagoudakis08,roumpos09}, and 
elementary excitations \cite{utsunomiya08}. If the trend followed by atom optics physics community holds for the exciton-polariton community,
one important branch of study of exciton-polariton BECs will be the application of periodic potentials on the BEC system. Optical lattices
have attracted much attention, spurred on by the experiments demonstrating the phase transition 
between a BEC and a Mott insulating state in a Bose-Hubbard model 
\cite{greiner02}. The application of the periodic potential simultaneously increases the particle-particle
interaction, as well as decreasing the kinetic energy. This allows the ratio of the Hubbard on-site interaction to the hopping amplitude $ U/t $ 
to be varied at will. 
The experiment has been of particular interest in the quantum information community, since the experiment realizes a nearly 
ideal quantum simulator \cite{feynman82,buluta09}. A quantum simulator is a device that directly recreates a quantum many-body problem in the laboratory. 
By experimentally modifying physical parameters, such as the periodic potential amplitude, temperature, and density, one may explore the phase
diagram of the system. 

\begin{figure}
\scalebox{0.6}{\includegraphics[bb=0 0 544 420]{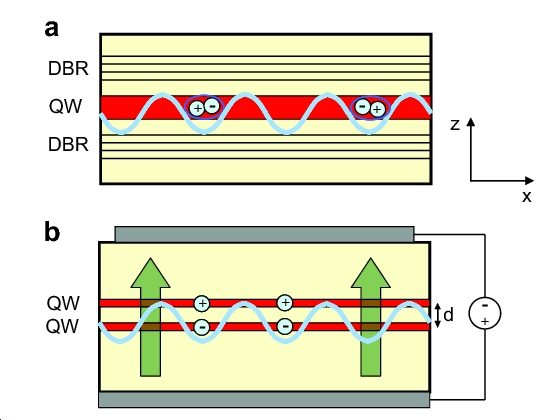}}
\caption{\label{fig:device} (Color online) 
Schematic device configurations considered in this paper. (a) Exciton-polaritons and (b) indirect excitons 
in a periodic potential.  A dc voltage is 
applied on the indirect excitons such that the holes and electrons occupy the top and bottom quantum wells (QW) respectively. Exciton-polaritons are formed by coupling the excitons to a distributed Bragg reflector (DBR). }
\end{figure}

The formation of a Bose-Hubbard model using polaritonic systems was first proposed in refs. \cite{hartmann06,greentree06}. In this paper we develop the theory for exciton-polaritons subject to a periodic potential (see Fig. \ref{fig:device}a). In contrast to the works of refs. \cite{hartmann06,greentree06,na08,aichorn08,koch09,grochol09}, where the 
polariton interaction originates from an effective nonlinearity due to a coupling to atomic sites, our interaction originates from the excitonic components of the polaritons, which ultimately originates
from a Coulomb interaction. Starting from a microscopic Hamiltonian for electrons and holes and their Coulomb interaction, we derive the origin of the Bose-Hubbard model that is assumed in Ref. \cite{na09}, allowing an accurate
determination of the Bose-Hubbard parameters $ U $ and $ t $.
From an experimental point of view, steps towards a similar experimental configuration as the optical lattice have been realized already by modifying the semiconductor microcavity system. In ref. \cite{lai07}, it was shown that a band structure was successfully formed using a metal deposition technique. The periodic metal structure on the surface changes the boundary conditions of the photon field, thus creating a static periodic potential for the polaritons \cite{kim08}.  Other methods for trapping polaritons have been proposed by etching the microcavity \cite{daif06}.  Such etching techniques are anticipated to produce stronger trapping potentials and access a more strongly correlated regime \cite{na08}.

The formal similarity in the treatment of 
indirect excitons allows us to write general formulas that capture both the polariton and indirect exciton interaction (see Fig. \ref{fig:device}b). Polaritons are described in the 
$ d = 0 $ limit of the formulas, where $ d $ is the bilayer separation of the indirect exciton system.  Indirect excitons have a non-zero $ d $, 
but a vanishing photon component. Although only exciton-polaritons have currently been observed to undergo Bose-Einstein condensation so far  \cite{keldysh95,butov02}, 
there is a large amount of interest in BECs of indirect excitons, as well as works of indirect excitons in periodic lattices \cite{remeika09}, motivating us to write the generalized formulas for both cases. We place particular interest on what parameters are required for observing a Bose-Hubbard Mott transition. For indirect excitons, as the bilayer separation $ d $ is increased, the bosonic nature of the excitons gradually diminishes due to 
the reduced electron-hole interaction.  The system is more appropriately described as a Fermi-Hubbard model in this limit.  We discuss the criterion for this crossover to occur for our model.  We also 
conjecture that a crossover between a Bose Mott insulator to a Fermi Mott insulator should take place 
with increasing bilayer separation, in analogy to the more commonly known BCS-BEC crossover \cite{comte82,littlewood96}.   
 
SI units are used throughout this paper.

\section{Bose-Hubbard model}
\label{sec:periodic}

We assume that a periodic potential of the form 
\begin{align}
\label{effectivepotential}
W_{\mbox{\tiny  ph}} (\bm{r}) & = W_0^{\mbox{\tiny  ph}}  \left[ \cos (k_0 x) + \cos (k_0 y) \right] \nonumber \\
W_{\mbox{\tiny  exc}} (\bm{r}) & = W_0^{\mbox{\tiny  exc}}  \left[ \cos (k_0 x) + \cos (k_0 y) \right] 
\end{align}
is applied on the photonic (ph) and excitonic (exc) components of the exciton-polaritons respectively, where $ k_0 = 2 \pi/ \lambda $ and $ \lambda $ is the wavelength of the periodic potential created. As mentioned in the introduction, a variety of experimental 
methods exist to create such a potential on the photon field \cite{rapaport05,lai07,kim08,daif06}. For the 
excitonic part, metal gates may be applied to the surface, trapping the excitons under the gates \cite{hammack06}. The potential $ W_{\mbox{\tiny  exc}} (\bm{r}) $ is an effective potential for the center of mass motion of the exciton, obtained after integrating over the relative motion of the exciton. The potential can in principle be either type I (where the electron and holes share the same potential minimum locations) or type II (electrons and holes have minima on alternate sublattices). For example, deformation potentials induced by surface acoustic waves \cite{rudolph07} and dipolar traps \cite{hammack06} are type I potentials. 
Meanwhile, the piezoelectric trapping technique \cite{lima05} is an example of a type II trapping potential. 
Type II potentials need rather strong trapping potentials for each individual component of the exciton (electron and hole) since their effective amplitude is suppressed by a factor of $ (k_0 a_B)^2 $ (see. eq. (30) in Ref. \cite{lima05}). However, since for strong potentials type II potentials tend to ionize the excitons, we believe that a type I potential is more promising in order to avoid these undesired effects.

The total Hamiltonian of the system is then
\begin{align}
\label{grandham}
H  & = H_{\mbox{\tiny exc}} + H_{\mbox{\tiny exc-exc}} + H_{\mbox{\tiny ph}} + H_{\mbox{\tiny exc-ph}} + H_{\mbox{\tiny sat}} \\
H_{\mbox{\tiny exc}}  & = \int d^2 r  b^\dagger (\bm{r}) \left[ - \frac{\hbar^2 }{2 M} \nabla^2  + W_{\mbox{\tiny  exc}} (\bm{r}) \right] b (\bm{r}) \\
H_{\mbox{\tiny exc-exc}} & = \frac{1}{2} \int d^2 Q d^2 Q' d^2 q
b^\dagger_{\bm{Q}-\bm{q}} b^\dagger_{\bm{Q}'+\bm{q}} U_{\mbox{\tiny exc}}(\bm{Q},\bm{Q}',\bm{q})
b_{\bm{Q}'}  b_{\bm{Q}} , \\
H_{\mbox{\tiny ph}}  & = \int d^2 r  a^\dagger (\bm{r}) \left[ - \frac{\hbar^2 }{2 m_{\mbox{\tiny ph}} } \nabla^2   + W_{\mbox{\tiny  ph}} (\bm{r}) \right] a (\bm{r}) \\
H_{\mbox{\tiny exc-ph}}  & = \hbar g \int d^2 q   \left[ b_{\bm{q}}^\dagger a_{\bm{q}} +
a_{\bm{q}}^\dagger b_{\bm{q}} \right] \\
H_{\mbox{\tiny sat}}  & = \frac{1}{2} \int d^2 Q d^2 Q' d^2 q \left[
a_{\bm{q}}^\dagger  b^\dagger_{\bm{Q}+\bm{Q}'-\bm{q}} U_{\mbox{\tiny sat}} (\bm{Q},\bm{Q}',\bm{q}) 
b_{\bm{Q}'}  b_{\bm{Q}} + \mbox{H.c.} \right], 
\end{align}
where
\begin{align}
\label{excitonicoperator}
b (\bm{r}) & = \frac{1}{2 \pi } \int d^2 q e^{i \bm{k} \cdot \bm{r} } b_{\bm{q}}  \\
\label{photonicoperator}
a (\bm{r}) & = \frac{1}{2 \pi } \int d^2 q e^{i \bm{q} \cdot \bm{r} } a_{\bm{q}}
\end{align}
are the annihilation operators for the quantum well excitons and microcavity photons respectively, $ M = m_e+m_h $ is the exciton mass, $ g $ is the exciton-photon coupling, and $ U_{\mbox{\tiny exc}} (\bm{r}, \bm{r}') $ is the effective interaction between two excitons. The photon acquires an effective mass 
$ m_{\mbox{\tiny ph}} $ through the 
dispersion in a two-dimensional microcavity, where the photon energy is $ E_{\mbox{\tiny ph}} = m_{\mbox{\tiny ph}} c^2 $, and $ c $ is the speed of light in GaAs. 
We do not consider the spin of the excitons explicitly because we assume that the polaritons are injected with a linear polarization such that only one spin species is present. The form of the exciton interaction is discussed in detail in section \ref{sec:interaction}.  $ H_{\mbox{\tiny sat}} $ is the non-linear interaction due to the exciton-photon coupling \cite{rochat00} and is discussed in section \ref{sec:saturation}. 

Substituting eqs. (\ref{excitonicoperator}) and (\ref{photonicoperator}) into (\ref{grandham}), and defining the upper ($ \sigma = \uparrow$) and lower ($ \sigma = \downarrow$) polariton 
operators
\begin{equation}
p_{\bm{q}}^\sigma = u^\sigma b_{\bm{q}} + v^\sigma a_{\bm{q}},
\end{equation}
where $ |u^\sigma |^2 + | v^\sigma |^2 = 1 $. We obtain the polariton Hamiltonian
\begin{align}
H & = H_{\mbox{\tiny kin}} + H_{\mbox{\tiny pot}} + H_{\mbox{\tiny pol}} \\
H_{\mbox{\tiny kin}} & = \int d^2 q \epsilon_{\bm{q}} p_{\bm{q}}^\dagger p_{\bm{q}} \label{polkineticenergy} \\
H_{\mbox{\tiny pot}} & = \frac{(| u |^2 W_0^{\mbox{\tiny  exc}}  +  | v |^2 W_0^{\mbox{\tiny  ph}})}{2} \int d^2 q 
\left[ p_{\bm{q}+ (k_0,0)}^\dagger p_{\bm{q}} + 
p_{\bm{q}+ (0,k_0)}^\dagger p_{\bm{q}} + \mbox{H.c.} \right] \\
H_{\mbox{\tiny pol}} & = \int d^2 Q d^2 Q' d^2 q
U_{\mbox{\tiny pol}}(\bm{Q},\bm{Q}',\bm{q}) p_{\bm{Q}-\bm{q}}^\dagger p_{\bm{Q}'+\bm{q}}^\dagger 
p_{\bm{Q}'} p_{\bm{Q}} \\
U_{\mbox{\tiny pol}} (\bm{Q},\bm{Q}',\bm{q}) & = \frac{| u |^4 }{2} U_{\mbox{\tiny exc}}(\bm{Q},\bm{Q}',\bm{q}) - \frac{| u |^2 (u v^*+u^* v)}{2} U_{\mbox{\tiny sat}}(\bm{Q},\bm{Q}',\bm{Q}-\bm{q}),
\end{align}
where we have only included terms where lower polariton operators appear, and compacted the 
notation such that $ p_{\bm{q}} \equiv p_{\bm{q}}^\downarrow $. Physically, disregarding the upper polariton operators corresponds to a low-temperature regime where there is negligible upper polariton population, which is routinely achieved experimentally. The Hopfield coefficients 
$ u \equiv u^\downarrow $ and $ v \equiv v^\downarrow $ are taken around $ \bm{q} = 0 $, again 
assuming that only the low energy states are excited. The lower polariton dispersion is obtained by expanding around 
$ \bm{q} = 0 $ giving
\begin{equation}
\epsilon_{\bm{q}} \approx \frac{\hbar^2 q^2}{2 m_{\mbox{\tiny pol}}} ,
\end{equation}
where the lower polariton mass is given by 
\begin{equation}
\frac{1}{m_{\mbox{\tiny pol}}} = \frac{|u |^2}{M} + \frac{| v |^2}{ m_{\mbox{\tiny ph}}}.
\end{equation}
Reverting to real space  makes it clear that we have polaritons in a periodic potential
\begin{equation}
\label{polaritonfinalham}
H = \int d^2 r  p^\dagger (\bm{r}) \left[ - \frac{\hbar^2 }{2 m_{\mbox{\tiny pol}}} \nabla^2  
+ | v |^2 W_{\mbox{\tiny  ph}} (\bm{r}) +| u |^2 W_{\mbox{\tiny  exc}} (\bm{r}) \right] p (\bm{r}) + H_{\mbox{\tiny pol}} .
\end{equation}

For sufficiently low temperatures, we may retain the lowest energy band of the Hamiltonian (\ref{polaritonfinalham}) to a good approximation.
A necessary temperature criterion is that the thermal energy $ k_B T $ is less than the band gap $\Delta $. In one dimension, any non-zero potential $ W_0 $ will open a bandgap. In two
dimensions however, for small potentials the lowest energy band overlaps in energy with the second lowest energy 
band. To ensure that these bands are separated, a potential of approximately 
$ W_0 \approx \frac{\hbar^2 k_0^2}{2 m_{\mbox{\tiny pol}}} $ is needed \cite{byrnes07},
where 
\begin{equation}
W_0 = | v |^2  W_0^{\mbox{\tiny ph}} + | u |^2  W_0^{\mbox{\tiny exc}}
\end{equation}
is the total potential amplitude due to exciton and photon parts. Under these circumstances, we may make a Wannier transformation and retain only states in the lowest energy band. This yields
\begin{equation}
\label{bosehubbardham}
H = \sum_{\bm{n},\bm{n}'} t(\bm{n},\bm{n}') p^\dagger_{\bm{n}} 
p_{\bm{n}'} + \frac{1}{2} \sum_{\bm{n}_1,\bm{n}_2,\bm{n}_3,\bm{n}_4} U( \bm{n}_1,\bm{n}_2,\bm{n}_3,\bm{n}_4) 
p^\dagger_{\bm{n}_1} p^\dagger_{\bm{n}_2} p_{\bm{n}_3} p_{\bm{n}_4}
\end{equation}
where
\begin{equation}
t(\bm{n},\bm{n}') = \int d^2 r w^* ( \bm{r} - \bm{n}\lambda ) \left[ - \frac{\hbar^2 }{2 m_{\mbox{\tiny pol}}} \nabla^2 +  | v |^2 W_{\mbox{\tiny ph}} (\bm{r}) 
+| u |^2 W_{\mbox{\tiny exc}} (\bm{r})  \right] 
w( \bm{r} - \bm{n}'\lambda ),
\label{tdefinition}
\end{equation}
\begin{eqnarray}
U( \bm{n}_1,\bm{n}_2,\bm{n}_3,\bm{n}_4) & = & \int d^2 r d^2 r' 
w^* ( \bm{r} - \bm{n}_1\lambda ) w^* ( \bm{r}' - \bm{n}_2\lambda )  U_{\mbox{\tiny pol}} (\bm{r},\bm{r}')  \nonumber \\
& & \times w ( \bm{r}' - \bm{n}_3\lambda ) w ( \bm{r} - \bm{n}_4\lambda ) ,
\label{coulombdefinition}
\end{eqnarray}
and
\begin{equation}
p_{\bm{n}} = \int d^2 r w ( \bm{r} - \bm{n}\lambda ) p ( \bm{r}) .
\end{equation}
$ w ( \bm{r} - \bm{n} \lambda) $ is the Wannier function centered around the lattice point $ \bm{n} = (n_x,n_y) $. 
The Hamiltonian (\ref{bosehubbardham}) is a Bose-Hubbard Hamiltonian.  We shall be only concerned with the nearest neighbor
tunneling matrix elements $ t $ and the on-site Coulomb interaction $ U $ in this paper:
\begin{equation}
\label{tnndefinition}
t \equiv t(\bm{n},\bm{n}+(1,0) ) = \int dx w^* ( x ) \left[ - \frac{\hbar^2 }{2 m_{\mbox{\tiny pol}}} \nabla^2 +  W_0 \cos (k_0 x) \right] 
w( x -\lambda ) ,
\end{equation}
and
\begin{eqnarray}
\label{udefinition}
U \equiv U( \bm{n},\bm{n},\bm{n},\bm{n}) & = &  \int d^2 r d^2 r' 
| w ( \bm{r} ) |^2   U_{\mbox{\tiny pol}} (\bm{r},\bm{r}')  | w ( \bm{r}' ) |^2  .
\end{eqnarray}
In writing (\ref{bosehubbardham}) we assume that the polariton lifetime $ \tau_{\mbox{\tiny}} $ is sufficiently long such that the extended ``superfluid'' and Mott states can occur. For example, for an extended ``superfluid'' state we require that there is enough time for the polariton to hop several times before decaying
\begin{equation}
\label{superlifetime}
\tau \gtrsim \frac{\hbar}{t} 
\end{equation}
Similarly for the Mott insulating state, the Coulomb energy should obey
\begin{equation}
\label{motttime}
\tau \gtrsim \frac{\hbar}{U} .
\end{equation}
In addition to the bandgap criterion $ k_B T < \Delta $, it is also necessary to have $ k_B T < U, t $,
such that only the low energy physics of the Bose-Hubbard model is probed. To evaluate (\ref{udefinition}) we require a form for the polariton-polariton interaction which we evaluate in the next section.

\section{Effective polariton interaction}

\subsection{Exciton-exciton interaction contribution}
\label{sec:interaction}

To obtain the effective interaction between excitons, we use the methods of de-Leon and Laikhtman \cite{deleon01}. 
There are many approaches to find the effective interaction between excitons in the literature, such as usage of the Usui transformation \cite{hanamura79}, variational wavefunction methods \cite{stolz81}, and operator methods \cite{bobrysheva72}. We find that the wavefunction methods of Ref. \cite{deleon01} are most transparent and systematically give the exciton interactions to order $ a_B^2/A $, where $ a_B = 4 \pi \epsilon \hbar^2 / 2 e^2 \mu $
is the 2D Bohr radius, $ A $ is the trapping area of the excitons, and $ \mu $ is the reduced mass
$ \mu = m_e m_h /(m_e + m_h) $. We note that a similar
method was used by  Ciuti {\it et al.} \cite{ciuti98} to find the same result for the Coulomb interactions, 
but it is unclear how to treat corrections to the kinetic energy operator (``kinematic corrections'') based 
solely on their method. Ref. \cite{deleon01} makes it clear that such terms cancel in the end and 
do not give rise to a physical interaction.  Here, we generalize the results to indirect excitons in a periodic potential. 

Following Ref. \cite{deleon01}, the effective Hamiltonian for the 
excitons can be decomposed into the following terms (omitting function labels for brevity):
\begin{align}
U_{\mbox{\tiny exc}}  = & U_{\mbox{\tiny dir}}  + U_{\mbox{\tiny exch}}^X 
+ U_{\mbox{\tiny exch}}^e  + U_{\mbox{\tiny exch}}^h  -\frac{1}{2}( {\cal A} {\cal H}_0 + {\cal H}_0 {\cal A}) + {\cal H}^{(exex)},
\label{deleonham}
\end{align}
where the terms are the direct exciton scattering, the exciton exchange scattering, the electron exchange 
scattering, the hole exchange scattering, the correction due to non-orthonormality of the wavefunctions, and
the contribution due to excited states of the excitons. Since only exciton-exciton interactions to order $ a_B^2/A $ are kept, 
only two-body exciton interactions need to be considered.  Explicit expressions for the above terms are as follows.  The direct term is 
%
\begin{equation}
U_{\mbox{\tiny dir}} (\bm{Q}, \bm{Q}',\bm{q}) = \int d^2 r_e d^2 r_h d^2 r_{e'} d^2 r_{h'}
\Psi_{\bm{Q}}^* (\bm{r}_e, \bm{r}_h) \Psi_{\bm{Q}'}^* (\bm{r}_{e'}, \bm{r}_{h'}) {\cal H} \Psi_{\bm{Q}+\bm{q}} (\bm{r}_e, \bm{r}_h) \Psi_{\bm{Q}'-\bm{q}} (\bm{r}_{e'}, \bm{r}_{h'}),
\label{directtermdefinition}
\end{equation}
where Hamiltonian for the two exciton system is
\begin{align}
{\cal H} & = {\cal H}_0 + {\cal H}_1 \nonumber \\
{\cal H}_0 & = -\frac{\hbar^2}{2m_e} \nabla_e^2 -\frac{\hbar^2}{2m_h} \nabla_h^2 -\frac{\hbar^2}{2m_{e'}} \nabla_{e'}^2 -\frac{\hbar^2}{2m_{h'}} \nabla_{h'}^2 -V( | \bm{r}_e - \bm{r}_h|) -V( | \bm{r}_{e'} - \bm{r}_{h'}|) \nonumber \\
{\cal H}_1 & = V( | \bm{r}_e - \bm{r}_{e'}|) + V( | \bm{r}_h - \bm{r}_{h'}|) - 
V( \sqrt{| \bm{r}_e - \bm{r}_{h'}|^2 +d^2}) - V(\sqrt{ | \bm{r}_h - \bm{r}_{e'}|^2 + d^2})
\end{align}
with $ V(r) = e^2/ 4 \pi \epsilon r $ ($ \epsilon \approx 13 \epsilon_0 $ is the 
permittivity in GaAs). The exciton exchange term is 
\begin{equation}
U_{\mbox{\tiny exch}}^X (\bm{Q}, \bm{Q}',\bm{q}) = U_{\mbox{\tiny dir}} (\bm{Q}, \bm{Q}',\bm{Q}' - \bm{Q} - \bm{q}) .
\end{equation}
The electron (hole) exchange terms are obtained by multiplying (\ref{directtermdefinition}) by $ -1$ and exchanging 
$ \bm{r}_e \leftrightarrow \bm{r}_e' $ ($ \bm{r}_h \leftrightarrow \bm{r}_h' $) in the final states. The 
correction due to non-orthonormality is
\begin{equation}
{\cal A}(\bm{Q}, \bm{Q}',\bm{q}) = -I_A^e(\bm{Q}, \bm{Q}',\bm{q}) -I_A^h(\bm{Q}, \bm{Q}',\bm{q})
\end{equation}
where
\begin{align}
I_A^e(\bm{Q}, \bm{Q}',\bm{q})& = \int d^2 r_e d^2 r_h d^2 r_{e'} d^2 r_{h'}
\Psi_{\bm{Q}}^* (\bm{r}_{e'}, \bm{r}_h) \Psi_{\bm{Q}'}^* (\bm{r}_{e}, \bm{r}_{h'} ) \Psi_{\bm{Q}+\bm{q}} (\bm{r}_{e}, \bm{r}_h) \Psi_{\bm{Q}'-\bm{q}} (\bm{r}_{e'}, \bm{r}_{h'}) \nonumber \\
I_A^h(\bm{Q}, \bm{Q}',\bm{q})& = \int d^2 r_e d^2 r_h d^2 r_{e'} d^2 r_{h'}
\Psi_{\bm{Q}}^* (\bm{r}_{e}, \bm{r}_{h'}) \Psi_{\bm{Q}'}^* (\bm{r}_{e'}, \bm{r}_{h} ) \Psi_{\bm{Q}+\bm{q}} (\bm{r}_{e}, \bm{r}_h) \Psi_{\bm{Q}'-\bm{q}} (\bm{r}_{e'}, \bm{r}_{h'}) .
\end{align}
The final term $ {\cal H}^{(exex)}$ originating from excited states of the excitons is neglected in our analysis since its magnitude is small, 
using similar arguments as given in Ref. \cite{deleon01}.  

To evaluate the expressions above, we use an approximate form for the indirect exciton ground state wavefunction, as obtained in ref. \cite{leavitt90}:
%
\begin{equation}
\Psi_{\bm{Q}} (\bm{r}_e, \bm{r}_h) = \frac{1}{\sqrt{A}} 
\exp [ i \bm{Q} \cdot (\beta_e \bm{r}_e + \beta_h \bm{r}_h) ]G(\rho,Z) f^{(e)} (z_e) f^{(h)} (z_h) ,
\label{approximateexciton}
\end{equation}
where $ \rho = \sqrt{ (x_e-x_h)^2 + (y_e-y_h)^2 } $, $ Z = z_e-z_h $, and $ \beta_{e,h} = m_{e,h}/(m_e + m_h) $. 
In eq. (\ref{approximateexciton}), the exciton is considered to be trapped in a large area $ A $, such that the center of mass wavefunction is of the form of a plane wave. 
The relative wavefunction of the exciton is
%
\begin{equation}
\label{excitonwavefunc}
G(\rho,Z) = \frac{N_G}{a_B} \exp \left[ - (\lambda(Z)/2) ( \sqrt{ (\rho/a_B)^2 + (Z/a_B)^2 } - Z/a_B ) \right] ,
\end{equation}
where $ \lambda(Z) = 2/(1+\sqrt{2 Z/a_B}) $ and $ N_G $ is a normalization factor such that 
$ \int |\Psi_{\bm{Q}} (\bm{r}_e, \bm{r}_h) |^2 d^3 r_e d^3 r_h = 1 $. $f^{(e)} (z) $ ($f^{(h)} (z) $) is the electron (hole) 
wavefunction satisfying the Schr{\"o}dinger equation in the $z$-direction. For indirect excitons with the electrons and holes confined to their 
respective quantum wells, we may approximate the wavefunctions to be $ f^{(e)} (z_e) = \delta (z_e - d/2) $,
$ f^{(h)} (z_h) = \delta (z_h + d/2) $, and $ Z= d $. The exciton wavefunction for polaritons may be recovered by setting $ d = 0 $. 

\begin{figure}
\scalebox{0.6}{\includegraphics[bb=0 0 488 288]{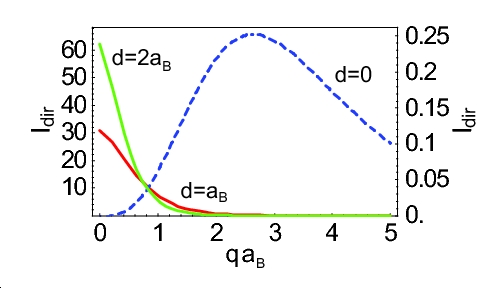}}
\caption{\label{fig:direct} (Color online) 
The direct exciton-exciton interaction integral $ I_{\mbox{\tiny dir}} (q,d) $ for three quantum well separations $ d/a_B = 0 $ 
(dashed line, right axis) and $ d/a_B = 1,2 $ (solid lines, left axis).   }
\end{figure}

\begin{figure}
\scalebox{0.6}{\includegraphics[bb=0 0 442 264]{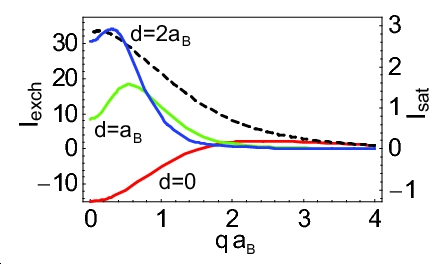}}
\caption{\label{fig:exchange} (Color online) 
The momentum transfer dependence of the electron-electron exchange  interaction integral 
$ I_{\mbox{\tiny exch}} ( \Delta Q=0,q,\theta,\beta_e,d)$ for $ d/a_B = 0,1,2 $ (solid lines, left axis) and all $ \theta $. 
The photon momentum  dependence of the saturation interaction integral 
$ I_{\mbox{\tiny sat}} (0,0,q)$ for $ d/a_B = 0 $ (dashed line, right axis). 
 }
\end{figure}

The evaluations of the direct and exchange terms are deferred to Appendix \ref{app:effectiveinteraction}. We find the direct term to be 
\begin{equation}
\label{finaldirect}
U_{\mbox{\tiny dir}} (\bm{Q}, \bm{Q}',\bm{q}) = \left[ -2E_{1s} + \frac{\hbar^2}{2M} (\bm{Q}^2 + \bm{Q}'^2) 
\right]\delta( \bm{q})
+\frac{1}{A} \frac{e^2}{4 \pi \epsilon} a_B \left( \frac{2}{\pi} \right)^2 I_{\mbox{\tiny dir}} (q,d) .
\end{equation}
where $ E_{1s} $ is the binding energy of a $1s$ exciton. The function $ I_{\mbox{\tiny dir}} (q,d) $ is plotted for various $ d $ in Fig. \ref{fig:direct}. 
The exciton exchange term is
\begin{align}
\label{excitonexchangeterm}
U_{\mbox{\tiny exch}}^X (\bm{Q}, \bm{Q}',\bm{q}) & = \left[ -2E_{1s} + \frac{\hbar^2}{2M} (\bm{Q}^2 + \bm{Q}'^2) 
\right]\delta( \bm{Q} - \bm{Q}' + \bm{q}) \nonumber \\
& + \frac{1}{A} \frac{e^2}{4 \pi \epsilon} a_B \left( \frac{2}{\pi} \right)^2  I_{\mbox{\tiny dir}} (\sqrt{(\Delta Q)^2 + q^2 - 2 \Delta Q q \cos \theta } ,d),
\end{align}
where $ \Delta Q = |\bm{Q}'- \bm{Q}| $ and $ \theta $ is the angle between $ \bm{Q}'- \bm{Q} $ and $ \bm{q} $. 
The electron and hole exchange terms are
\begin{align}
\label{electronexchange}
U_{\mbox{\tiny exch}}^e (\bm{Q}, \bm{Q}',\bm{q}) & = 
-\left[ -2E_{1s} + \frac{\hbar^2}{4M} (\bm{Q}^2 + \bm{Q}'^2) + \frac{\hbar^2}{4M} ((\bm{Q}+\bm{q})^2 + (\bm{Q}'-\bm{q})^2) 
\right] I_A^e(\bm{Q}, \bm{Q}',\bm{q}) \nonumber \\
& - \frac{1}{A} \frac{e^2}{4 \pi \epsilon} 
a_B \left( \frac{2}{\pi} \right)^2  I_{\mbox{\tiny exch}} ( \Delta Q,q,\theta,\beta_e,d)  \\
\label{holeexchange}
U_{\mbox{\tiny exch}}^h (\bm{Q}, \bm{Q}',\bm{q}) & =  
-\left[ -2E_{1s} + \frac{\hbar^2}{4M} (\bm{Q}^2 + \bm{Q}'^2) + \frac{\hbar^2}{4M} ((\bm{Q}+\bm{q})^2 + (\bm{Q}'-\bm{q})^2) 
\right] I_A^h(\bm{Q}, \bm{Q}',\bm{q}) \nonumber \\
& - \frac{1}{A} \frac{e^2}{4 \pi \epsilon} 
a_B \left( \frac{2}{\pi} \right)^2  I_{\mbox{\tiny exch}} ( \Delta Q,q,\theta,\beta_h,d) ,
\end{align}
where in the square brackets we evaluated half of the operator on the initial states and half on the final states. Doing this we see that these terms exactly cancel with the corrections due to non-orthonomality (i.e. the fifth term in (\ref{deleonham})). 
Numerical evaluations of the exchange integral $ I_{\mbox{\tiny exch}} $ are shown in Fig. \ref{fig:exchange}.

Substituting (\ref{finaldirect}-\ref{holeexchange}) into (\ref{deleonham}) we obtain the final effective Hamiltonian for the 
two-exciton system.  Subtracting the kinetic energy and binding energy terms, we obtain an expression for exciton-exciton interaction 
\begin{eqnarray}
U_{\mbox{\tiny exc}}(\bm{Q}, \bm{Q}',\bm{q}) & = &
\frac{1}{A} \frac{e^2}{4 \pi \epsilon} 
a_B \left( \frac{2}{\pi} \right)^2 \Big[ I_{\mbox{\tiny dir}} (q,d) + I_{\mbox{\tiny dir}} (\sqrt{(\Delta Q)^2 + q^2 - 2 \Delta Q q \cos \theta } ,d) \nonumber \\
& & - I_{\mbox{\tiny exch}} ( \Delta Q,q,\theta,\beta_e,d) - I_{\mbox{\tiny exch}} ( \Delta Q,q,\theta,\beta_h,d) \Big]
\label{totalinteractionfull}
\end{eqnarray}

\begin{figure}
\scalebox{0.6}{\includegraphics[bb=0 0 467 310]{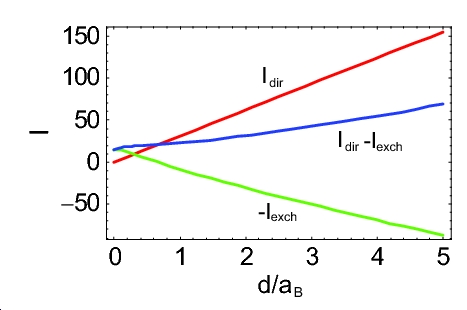}}
\caption{\label{fig:coulomb_d_dependence} (Color online) 
The direct Coulomb interaction $ I_{\mbox{\tiny dir}} (q=0,d) $, the negative of the 
electron-electron exchange interaction $ - I_{\mbox{\tiny exch}} ( \Delta Q=0,q=0,\theta,\beta,d)$,
and their sum versus the quantum well separation $ d $ for all $ \theta $ and $ \beta $.}
\end{figure}

The dependence on $ d $ for small $ q $ may be evaluated exactly for the direct term by 
expanding the term in the square brackets in (\ref{directtermone}). We obtain
%
\begin{equation}
U_{\mbox{\tiny dir}} (\bm{Q}, \bm{Q}',\bm{q}=0) =  \frac{d e^2}{ \epsilon A} .
\end{equation}
This is the zero momentum limit of the Fourier transform of the Coulomb interaction for oriented dipoles $ U(q) = e^2 (1- \exp(-dq))/\epsilon q$.  The $ d $ dependence of the exchange term
must be evaluated numerically, and our results are shown in Fig. \ref{fig:coulomb_d_dependence}.  
We find an approximately linear dependence of the exchange term with $ d $, which changes sign at $ d/a_B \approx 0.66 $. In Fig. \ref{fig:coulomb_d_dependence} we also plot the combined contributions of the direct and exchange integrals. We see that the total interaction remains repulsive for all $ d $, despite the exchange term changing sign.

\subsection{Saturation contribution}
\label{sec:saturation}

The saturation contribution to the polariton interaction comes from the coupling of the electron and holes to the electromagnetic field
\begin{align}
H_{\mbox{\tiny EM}} = \hbar {\cal G} \int d^2 k d^2 k' \left[ e^\dagger_{\bm{k}} h^\dagger_{\bm{k}'} a_{\bm{k}+\bm{k}'} + a_{\bm{k}+\bm{k}'}^\dagger h_{\bm{k}'} e_{\bm{k}}  \right] .
\label{emham}
\end{align}
The last two terms in (\ref{grandham}) may be found by considering the matrix element between the states
\begin{equation}
|\bm{Q}, \bm{Q}' \rangle = \frac{1}{2} 
\int d^2 r_e d^2 r_h d^2 r_{e'} d^2 r_{h'}
\Phi_{\bm{Q} \bm{Q}'} (\bm{r}_e, \bm{r}_h,\bm{r}_{e'}, \bm{r}_{h'} )
e^\dagger ( \bm{r}_e) h^\dagger (\bm{r}_h) e^\dagger (\bm{r}_{e'}) h^\dagger (\bm{r}_{h'}) |0 \rangle
\end{equation}
and
\begin{equation}
|\bm{Q}'', \bm{q} \rangle = 
a^\dagger_{\bm{q}} \int d^2 r_e d^2 r_h \Psi_{\bm{Q}''} (\bm{r}_e, \bm{r}_h)
e^\dagger ( \bm{r}_e) h^\dagger (\bm{r}_h) |0 \rangle
\end{equation}

where 
\begin{align}
\Phi_{\bm{Q} \bm{Q}'} (\bm{r}_e, \bm{r}_h,\bm{r}_{e'}, \bm{r}_{h'} ) 
& = \frac{1}{2} \Big[ \Psi_{\bm{Q}} (\bm{r}_e, \bm{r}_h) \Psi_{\bm{Q}'} (\bm{r}_{e'}, \bm{r}_{h'}) 
+ \Psi_{\bm{Q}} (\bm{r}_{e'}, \bm{r}_{h'}) \Psi_{\bm{Q}'} (\bm{r}_e, \bm{r}_h) \\
& - \Psi_{\bm{Q}} (\bm{r}_e, \bm{r}_{h'}) \Psi_{\bm{Q}'} (\bm{r}_{e'}, \bm{r}_{h}) 
- \Psi_{\bm{Q}} (\bm{r}_{e'}, \bm{r}_{h}) \Psi_{\bm{Q}'} (\bm{r}_{e}, \bm{r}_{h'}) \Big] .
\end{align}
Starting from the two exciton wavefunction, $ H_{\mbox{\tiny EM}}  $ can either destroy one of the excitons and produce a photon, or take an electron and hole
from each of the excitons and produce a photon.  These two processes give rise to the matrix element
\begin{align}
\langle \bm{Q}+\bm{Q}'-\bm{q}, \bm{q} | H_{\mbox{\tiny EM}} |\bm{Q}, \bm{Q}' \rangle =
\hbar {\cal G} \sqrt{A} G(0,0) ( \delta (\bm{Q}'-\bm{q}) 
+ \delta (\bm{Q}-\bm{q}) ) \nonumber \\
+  U_{\mbox{\tiny sat}} (\bm{Q},\bm{Q}',\bm{q}) + U_{\mbox{\tiny sat}} (\bm{Q}',\bm{Q},\bm{q}) ,
\label{saturationterm}
\end{align}
where
\begin{align}
\label{saturationintegral}
U_{\mbox{\tiny sat}} (\bm{Q},\bm{Q}',\bm{q}) = - \hbar {\cal G} \int d^2 x d^2 r_e d^2 r_h  \Psi_{\bm{Q}} (\bm{r}_{e}, \bm{x}) \Psi_{\bm{Q}'} (\bm{x},\bm{r}_{h}) 
\Psi_{\bm{Q}+\bm{Q}'-\bm{q}}^* (\bm{r}_{e}, \bm{r}_{h}) e^{-i\bm{q}\cdot \bm{x}} ,
\end{align}
and we have set $ \bm{Q}'' = \bm{Q} + \bm{Q}' - \bm{q} $ due to momentum conservation. 
The first two terms in (\ref{saturationterm}) correspond to the destruction of an exciton to 
create a photon, with another exciton acting as a bystander. This corresponds to the second last term in 
(\ref{grandham}). The last two terms correspond to an electron and 
hole being taken from each exciton, resulting in a new exciton being formed from the remaining electron and hole. This process clearly requires two excitons in the initial state, giving the non-linear last term in (\ref{grandham}). 
We only consider the case of zero bilayer separation ($d=0 $) here, since the Hamiltonian (\ref{emham}) requires that the electron and hole recombine into a photon at the same spatial position. In our approximation (\ref{approximateexciton}) where 
the electron and hole wavefunctions are perfectly confined to their respective quantum wells, for non-zero bilayer separation there is zero overlap of the 
electron and hole wavefunction, which gives a zero matrix element for (\ref{saturationterm}). 

We thus find
\begin{equation}
\label{rabisplitting}
g = {\cal G} \sqrt{A} G(0,0)
\end{equation}
and 
\begin{equation}
U_{\mbox{\tiny sat}} (\bm{Q},\bm{Q}',\bm{q}) = - \hbar {\cal G} \frac{a_B}{\sqrt{A}} I_{\mbox{\tiny sat}} (\bm{Q},\bm{Q}',\bm{q}) .
\label{saturationintegral2}
\end{equation}
where $ I_{\mbox{\tiny sat}} (\bm{Q},\bm{Q}',\bm{q}) $ is evaluated in the appendix. A numerical 
evaluation of the integral as a function of the photon momentum is shown in Fig. \ref{fig:exchange}.

\section{Mott-Hubbard transition}
\label{sec:motthubbard}

Returning to the effective Bose-Hubbard Hamiltonian of (\ref{bosehubbardham}), we may now estimate the size of the tunneling and Coulomb energies from
eqs.  (\ref{tnndefinition}) and (\ref{udefinition}). Starting from (\ref{udefinition}) we make a change of variables $ \bm{R}_{\mbox{\tiny CM}} =
\bm{R} + \bm{R}' $ and $ \bm{\rho} = (\bm{R} - \bm{R}')/2 $, giving
\begin{eqnarray}
U & = &  \int d^2 R_{\mbox{\tiny CM}} d^2 \rho 
|w_{\gamma} ( \bm{R}_{\mbox{\tiny CM}} + \bm{\rho}/2 )|^2 |w_{\gamma} ( \bm{R}_{\mbox{\tiny CM}} - \bm{\rho}/2 )|^2  U_{\mbox{\tiny pol}} (\bm{\rho}) .
\label{uintegralappd}
\end{eqnarray}
From Figs. \ref{fig:direct} and \ref{fig:exchange} we see that interaction is large up to a momentum of the order of $ \sim 1/a_B $. Thus 
the largest contributions to the above integral occur when the variable $ \rho  $ is of the order of $ \sim a_B $, which is much smaller than the length scale of the Wannier functions $ \sim \lambda $. We may thus write 
\begin{align}
U & \approx   \int d^2 \rho  U_{\mbox{\tiny pol}} (\bm{\rho}) \int d^2 R_{\mbox{\tiny CM}} 
|w_{\gamma} ( \bm{R}_{\mbox{\tiny CM}} )|^4  \nonumber \\
& =    U^{\mbox{\tiny coul}} + U^{\mbox{\tiny sat}}
\end{align}
where 
\footnote{The factor of $ A $ follows from the fact that the Fourier transform differs from the 
expression (\ref{totalinteractionfull}) by a factor of $ A $. All expressions for the effective interaction calculated here differ from the Fourier transformed interactions by a factor of $ A $, i.e. the expressions (\ref{finaldirect}-\ref{holeexchange})
have dimensions of energy, while Fourier transformed expressions have dimensions of
(energy)$\times$(area) (in 2D).}
\begin{align}
 U^{\mbox{\tiny coul}} & =  |u|^4 U_{\mbox{\tiny exc}} (0,0,0)  A \int d^2 R |w_{\gamma} (  \bm{R} )|^4 \nonumber \\
 U^{\mbox{\tiny sat}}  & = - |u|^2 (u v^* + u^* v) U_{\mbox{\tiny sat}} (0,0,0) A \int d^2 R |w_{\gamma} (  \bm{R} )|^4 .
\label{uexpressionsimple}
\end{align}

\begin{figure}
\scalebox{0.6}{\includegraphics[bb=0 0 540 601]{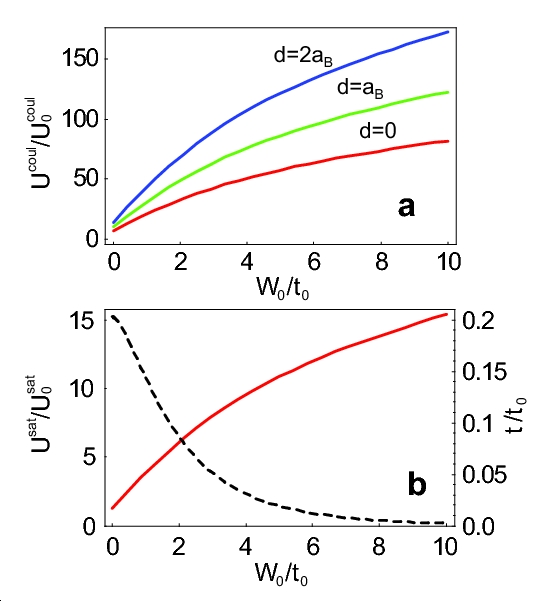}}
\caption{\label{fig:Uandtplot} (Color online) 
(a) Dependence of the on-site Coulomb energy $ U^{\mbox{\tiny coul}} $ for three bilayer separations $ d $ with the periodic potential amplitude $ W_0 $. Energies are normalized to units $ t_0 = \frac{h^2}{8 m_{\mbox{\tiny pol}} \lambda^2} $ and $ U_0^{\mbox{\tiny coul}} =\frac{2 e^2 a_B |u|^4}{\pi^3 \epsilon \lambda^2} $. 
(b) Dependence of the nearest neighbor hopping $ t $ (dashed line, right axis) and the on-site saturation energy $ U^{\mbox{\tiny sat}} $ (solid line, left axis) with the periodic potential amplitude $ W_0 $. The saturation energy scale is $ U_0^{\mbox{\tiny sat}} = 2 \hbar g \sqrt{\frac{\pi}{2}} |u|^2 \mbox{Re} (uv^*) \frac{a_B^2}{\lambda^2}  $. 
For example, for polaritons  in GaAs with $ \lambda = 0.5 $ $\mu$m, $ a_B = 10 $ nm, $2\hbar g = 15$ meV, and $ m_{\mbox{\tiny pol}} =0.01 M $ the energy scales are $t_0 = 0.91$meV,  $ U_0^{\mbox{\tiny coul}} = 3.58\mu$eV, and $ U_0^{\mbox{\tiny sat}} = 0.34\mu $eV. 
}
\end{figure}

Fig. \ref{fig:Uandtplot} shows the two contributions to $ U$ as well as the hopping amplitude $ t $.  The results are normalized to the characteristic
energies
\begin{align}
U_0^{\mbox{\tiny coul}} & = \frac{2 e^2 a_B |u|^4}{\pi^3 \epsilon \lambda^2} \\
U_0^{\mbox{\tiny sat}} & = 2 \hbar g \sqrt{\frac{\pi}{2}} |u|^2 \mbox{Re} (uv^*) \frac{a_B^2}{\lambda^2}   \\
t_0 & = \frac{h^2}{8 m_{\mbox{\tiny pol}} \lambda^2} .
\end{align}
where we have used (\ref{rabisplitting}) to convert $ \cal G $ into $ g $. Here, $ 2 \hbar g $ is the 
Rabi splitting of the polaritons. For polaritons, the bilayer separation is $ d = 0 $, while for indirect excitons the exciton component is $ | u |^2 = 1 $.  We see that increasing the potential strength $ W_0 $ decreases the hopping $ t $ while increasing 
$ U $, as expected. Increasing $ d $ enhances $ U^{\mbox{\tiny coul}} $, as the dipole moment of the 
excitons are enhanced with an increasing $ d $. 

We now derive a criterion
for a quantum phase transition from a BEC state into a Mott insulator state. In two dimensions, the phase transition is expected to occur at approximately  \cite{sheshadri93}
\begin{equation} 
\label{bhcriterion}
 U/t \approx 23 .
\end{equation}
By turning up the potential $ W_0 $ it is clear that at some point $ U/t $ will reach this critical amplitude.
The other variable that may be changed to reach the phase transition is the 
detuning of the polaritons which changes the polariton mass. For a potential of size $ W_0 \approx \frac{\hbar^2 k_0^2}{2 m_{\mbox{\tiny pol}}} $, we may derive a criterion for the polariton mass necessary to 
reach the phase transition using (\ref{bhcriterion}) and the ratio of the dimensionless parameters in Fig. \ref{fig:Uandtplot}. For GaAs, this gives
\begin{equation}
\label{masscriterion}
m_{\mbox{\tiny pol}} \approx 10^{-3} \frac{\pi^3 \epsilon h^2}{e^2 a_B | u|^4 }  \approx 10^{-2} M ,
\end{equation}
where in the second relation we assumed that $ | u|^2 $ is of the order of unity, and $ a_B \approx 10$ nm. This corresponds to extremely far blue-detuned polaritons, i.e. very exciton-like polaritons. The lack of the dependence on the wavelength $ \lambda $ is due to the 
cancellation of the dependence of the Coulomb interaction energy $ U_0 $ and the 
kinetic energy $ t_0 $.  The wavelength $ \lambda $ is still important however, 
as it sets the energy scale of the whole Bose-Hubbard Hamiltonian.  The energy scale should be set such that the parameters $ U $ and $ t $ 
are larger than the temperature of the experiment, such that only the lowest energy band is occupied. Furthermore, 
semiconductor systems possess an inevitable disorder potential due to reasons such as crystal imperfection and 
damage during fabrication, thus $ \lambda $ should be set small enough such that the hopping energy $ t $ overcomes this  disorder potential strength. 
For larger potentials than $ W_0 \approx \frac{\hbar^2 k_0^2}{2 m_{\mbox{\tiny pol}}} $, a lighter polariton mass is allowable for reaching the phase transition. Thus in practice a combination of blue-detuned polaritons and large potentials is probably the most favorable experimental configuration.

For example, using typical experimental parameters for polaritons ($d=0$) in GaAs using the criterion (\ref{masscriterion}), corresponding to $ u \approx 0.999 $, with  $ \lambda = 0.5 $ $\mu$m, $ a_B = 10 $ nm, $2\hbar g = 15$ meV, and an applied potential of $ W_0 = 6 $ meV, we obtain $ U = 0.24 $ meV and $ t = 9 $ $\mu$eV. This corresponds to temperatures in the vicinity of $ T \approx 0.1 $ K, which are reachable using today's refrigeration methods. In order that the system is stable in the Mott insulating state, the lifetime of the polaritons should be longer than the timescale set by (\ref{motttime}). Assuming that the lifetime of the very exciton-like polaritons can be approximated by typical exciton lifetimes $ \tau \approx 1 $ns \cite{feldmann87}, this corresponds to an energy scale $ U > \hbar/\tau \approx 1 \mu$eV. We thus see that for the above parameters the lifetime requirement is satisfied.

The Coulomb interaction is increased for indirect excitons as shown in Fig. \ref{fig:Uandtplot}. However, the increase is fairly modest for 
bilayer separations of the order of the Bohr radius. Thus considering that indirect excitons have not been observed to undergo BEC yet, 
the moderate advantage of increased interactions (at the sacrifice of a lighter polariton mass) is outweighed by the difficulty of cooling the system into the ground state. 

The state of the polaritons may be measured using standard photoluminescence measurements that measure the coherence across the condensate \cite{deng07}. In analogy to the 
experiments of Greiner {\it et al.} \cite{greiner02}, the transition to the Mott insulator state should lead to a lack of 
spatial coherence across the sample, resulting in the destruction of the far-field interference pattern. 
However, the disappearance of interference fringes does not unambiguously demonstrate the presence of a 
Mott insulator, since an uncondensed state will also have the same interference characteristics.  Therefore, 
a second order coherence Hanbury-Brown-Twiss measurement is also necessary to determine the correlations 
between the photons emanating from the sample.  At unit filling in a Mott insulator state, the conditional probability of detection of a photon originating from a particular site following detection from the same 
site vanishes.  A similar diminished probability is present at higher filling factors. This is the identical
 technique used to observe anti-bunching behavior in single photon generation.

\section{Fermi-Bose crossover boundary}
\label{sec:fermibose}

We now turn to the effect on the particle statistics of indirect excitons as the bilayer separation is increased. 
For $ d=0 $, excitons are well approximated by bosons for sufficiently low density, which motivated us to describe the system as a Bose-Hubbard model in section \ref{sec:periodic}.  In the limit $ d = \infty $, excitons cannot be described by bosons, and are more properly described as a double Fermi-Hubbard model, 
with electron-hole interactions between them. Since the two descriptions are rather different with 
differing phase transition criteria, it is of interest to know at what $ d $ this crossover occurs. 
The bosonic nature of (or the lack of) the excitons may be seen by examining the commutation relation
\begin{equation}
\label{bcommutation}
[ b , b^\dagger ] = 1 -K 
\end{equation}
where
\begin{equation}
\label{excitonoperator}
b^\dagger = \int d^2 r_e d^2 r_h \phi_{d} (\bm{r}_e - \bm{r}_h) w  ( \beta_e \bm{r}_e + \beta_h \bm{r}_h ) 
e^\dagger (\bm{r}_e) h^\dagger (\bm{r}_h)
\end{equation}
and
\begin{eqnarray}
K & = & \int d^2 r_e d^2 r_h d^2 r \phi_{d} ( \bm{r}_e - \bm{r}_h) w  ( \beta_e \bm{r}_e + \beta_h \bm{r}_h )
\Big[ \phi_{d} ( \bm{r}_e - \bm{r}) w  ( \beta_e \bm{r}_e + \beta_h \bm{r} ) 
h^\dagger (\bm{r}_h) h (\bm{r}) \nonumber \\
& & + \phi_{d} ( \bm{r}_h - \bm{r}) w  ( \beta_e \bm{r} + \beta_h \bm{r}_h ) 
e^\dagger (\bm{r}_e) e (\bm{r})
\Big] .
\label{kdefinition}
\end{eqnarray}
We have omitted the label $ \bm{n} $ and written $ b^\dagger_{\bm{n}} \rightarrow b^\dagger $ to simplify the notation. Only the commutation relation in the same potential minima of the periodic potential is considered here (i.e. the same Wannier function), as deviations from bosonic behavior should be most apparent in this case. The operator $ K $ may be interpreted as the correction operator to the commutation relation
(\ref{bcommutation}), as it contains the non-bosonic component of the exciton operators.

Due to the presence of the $ K $ operator, $ n $-particle states defined using the 
$ b^\dagger $ operators defined in (\ref{excitonoperator}) do not have the simple 
$ 1/\sqrt{n!} $ normalization of bosonic states. We must instead define such 
states according to 
\begin{equation}
\label{nstatedefinition}
| n \rangle = \frac{1}{\sqrt{N(n) n!} } (b^\dagger)^n | 0 \rangle,
\end{equation}
where $ N(n) $ is present for proper normalization. A derivation of this normalization 
factor is given in Appendix \ref{app:bosefermi}, up to 
powers linear in the correction operator $ K$. We obtain
\begin{equation}
\label{normalizationfactor}
N(n) = 1 - \frac{1}{4}n(n-1)  I_{\mbox{\tiny BF}} (W_0,d) ,
\end{equation}
where $ I_{\mbox{\tiny BF}} (W_0,d) $ is the integral expression given by (\ref{idefinition}) and has
an order of magnitude  $ \sim a_B^2 / \lambda^2 $. All terms neglected in (\ref{normalizationfactor}) have higher powers of $ a_B^2 / \lambda^2 $, which have a small contribution for the typical periodic potential dimensions that are 
possible using current fabrication methods ($\lambda \gg a_B $). 

\begin{figure}
\scalebox{0.6}{\includegraphics[bb=0 0 439 349]{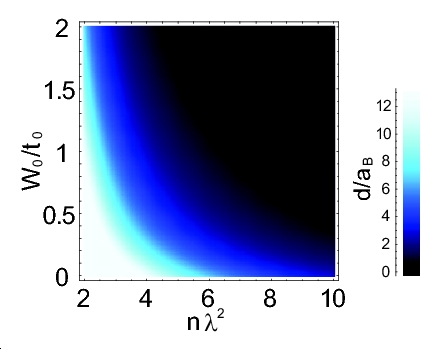}}
\caption{\label{fig:bose_fermi} 
(Color online) Boundary between Bose and Fermi behavior for the quantum well separation $ d $ (in units of the Bohr radius $ a_B $), 
the density $ n $ and the potential $ W_0  $. Well separations less than the indicated value indicate bosonic 
behavior, while Fermi behavior holds for the larger separations. ``Fermi'' behavior means that
the fermionic nature of the electron and hole making up the excitons become important.
 }
\end{figure}

Defined in the way (\ref{nstatedefinition}), the states $ | n \rangle $ provide an orthonormal basis set. Deviations from bosonic behavior occur due to the 
operator $ b^\dagger $ not providing the correct mapping between these states $ \langle n | \frac{b^\dagger}{\sqrt{n}} | n-1 \rangle \ne 1 $. Using the definition 
(\ref{nstatedefinition}), we have
\begin{equation}
\langle n | \frac{b^\dagger}{\sqrt{n} } | n-1 \rangle = \sqrt{\frac{N(n)}{N(n-1)}} .
\end{equation}
Deviations from unity of the RHS represents non-bosonic behavior. Substituting (\ref{normalizationfactor}), this factor is to 
lowest order in $  I_{\mbox{\tiny BF}} (W_0,d) $
\begin{equation}
\sqrt{\frac{N(n)}{N(n-1)}} \approx 1 - \frac{1}{2} (n-1)  I_{\mbox{\tiny BF}} (W_0,d) .
\end{equation}
Since the Bose to Fermi transition is a smooth crossover \cite{comte82}, strictly speaking it is arbitrary where to mark the boundary. However, a reasonable criterion for the 
location of the crossover from bosonic to fermionic behavior may be defined as when the second term 
in the above expression becomes of the order of unity:
\begin{equation}
\frac{1}{2} (n-1)  I_{\mbox{\tiny BF}} (W_0,d) = 1 .
\label{crossovercrit}
\end{equation}
The solution to the above criterion is plotted in  Fig. \ref{fig:bose_fermi}
in the space $ (d,W_0) $. We see that with decreasing $ n $, $ d $ and $ W_0 $, the excitons become more boson-like. This dependence on $ n $ and $ d $ is a restatement of the well-known result that the excitons become non-bosonic 
when their wavefunctions start overlapping, i.e. $ n a_B^2/ A \sim 1 $. The dependence on $ W_0 $ may 
be understood by considering the spread of the Wannier functions with $ W_0 $. As $ W_0 $ is 
increased, the Wannier functions become more localized, effectively reducing the area that 
the excitons are confined in. This enhances the overlap between the excitons, thus pushing the 
boundary towards the fermion side of the crossover. For $ n \le 1$, there is no solution to 
(\ref{crossovercrit}), meaning that the boundary for bosonic behavior extends all the way to infinity in $ W_0 $ and $ d $. The fact that solutions for bosonic behavior exist with $ n>1 $ means that in reality one cannot treat the excitons completely as hard- or soft-core bosons, and their true nature lies in between
these two limits.

\section{Fermionic description of the bilayer system}
\label{sec:fermionic}

\begin{figure}
\scalebox{0.6}{\includegraphics[bb=0 0 480 278]{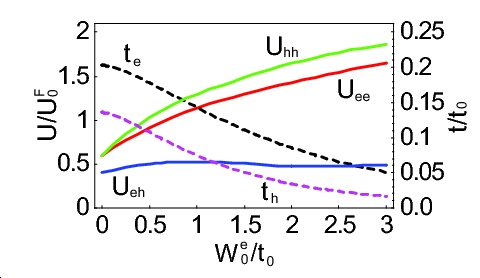}}
\caption{\label{fig:Uandtfermi} (Color online) 
Dependence of the nearest neighbor hopping for electrons $ t_e $ and holes $ t_h $ (dashed lines, right axis), as well as the electron-electron on-site Coulomb energy $ U_{ee} $, 
the hole-hole on-site Coulomb energy $ U_{hh} $, and the 
the electron-hole on-site Coulomb energy $ U_{eh} $ for $ \lambda \gg d $ (solid lines, left axis). 
Energies are normalized to units $ t_0 = \frac{h^2}{8m_e \lambda^2} $ and $ U_0^F =\frac{e^2 }{4 \epsilon \lambda} $. For example, for $ \lambda = 0.1 $ $\mu$m, $ a_B = 10 $ nm in GaAs, the energy scales are $t_0 = 0.561$meV and $ U_0^F = 3.48$meV. Parameters for GaAs are assumed here, with $ m_e= 0.067m_0 $, $ m_h= 0.1m_0 $, and $ \epsilon = 13 \epsilon_0 $, where $ m_0 $ is the free electron mass and $  \epsilon_0 $ is the permittivity of free space. }
\end{figure}

For parameter regions where the bosonic approximation is invalid, we must write the 
Hamiltonian in its full form involving both 
electron $ e_\sigma (\bm{r}) $ and hole $ h_\sigma (\bm{r}) $ operators:
\begin{align}
H  = & 
\sum_\sigma \int d^2 r  e^\dagger_\sigma (\bm{r}) \left[ \frac{\hbar^2 }{2 m_e} \nabla^2 + W_e (\bm{r}) \right] e_\sigma (\bm{r})  + \sum_\sigma \int d^2 r  h^\dagger_\sigma (\bm{r}) \left[ \frac{\hbar^2 }{2 m_h} \nabla^2 - W_e (\bm{r}) \right] h_\sigma (\bm{r}) \nonumber \\
 & + \frac{1}{2} \sum_{\sigma,\sigma'} \int \int d^2 r d^2 r' \Big[ 
e^\dagger_{\sigma'} (\bm{r}) e^\dagger_{\sigma} (\bm{r}') V ( | \bm{r} - \bm{r}' | )  e_{\sigma} (\bm{r}')  e_{\sigma'} (\bm{r} )
\nonumber \\
 & + h^\dagger_{\sigma'} (\bm{r}) h^\dagger_{\sigma} (\bm{r}') V ( | \bm{r} - \bm{r}' | )  
h_{\sigma} (\bm{r}')  h_{\sigma'} (\bm{r} )
-2 e^\dagger_{\sigma'} (\bm{r}) h^\dagger_{\sigma} (\bm{r}') V ( \sqrt{| \bm{r} - \bm{r}' |^2 + d^2})  h_{\sigma} (\bm{r}') e_{\sigma'} (\bm{r} ) \Big], \nonumber \\
\end{align}
where 
\begin{equation}
W_e (\bm{r}) = W_0^{\mbox{\tiny  e}}  \left[ \cos (k_0 x) + \cos (k_0 y) \right]  ,
\end{equation}
and we have assumed periodic potentials of equal magnitude, but opposite sign, are applied on the electron and hole.
Analogously to (\ref{bosehubbardham}),  we may transform to the Wannier
basis to give a electron-hole two-band Hubbard model \cite{ogawa05,tomio06}:
\begin{eqnarray}
H & = & \sum_\sigma \sum_{\bm{n},\bm{n}'} \Big[ t_e(\bm{n},\bm{n}') e^\dagger_{\bm{n}\sigma } e_{\bm{n}' \sigma}
+ t_h(\bm{n},\bm{n}') h^\dagger_{\bm{n} \sigma } h_{\bm{n}' \sigma}
 \Big] \nonumber \\
& & + \frac{1}{2} \sum_{\sigma,\sigma'} \sum_{\bm{n}_1,\bm{n}_2,\bm{n}_3,\bm{n}_4} \Big[ 
U_{ee} ( \bm{n}_1,\bm{n}_2,\bm{n}_3,\bm{n}_4) 
e^\dagger_{\bm{n}_1 \sigma'} e^\dagger_{\bm{n}_2 \sigma} e_{\bm{n}_3 \sigma} e_{\bm{n}_4 \sigma'} \nonumber \\
& & + U_{hh} ( \bm{n}_1,\bm{n}_2,\bm{n}_3,\bm{n}_4) 
h^\dagger_{\bm{n}_1 \sigma'} h^\dagger_{\bm{n}_2 \sigma} h_{\bm{n}_3 \sigma} h_{\bm{n}_4 \sigma'} 
- 2 U_{eh} ( \bm{n}_1,\bm{n}_2,\bm{n}_3,\bm{n}_4) 
e^\dagger_{\bm{n}_1 \sigma'} h^\dagger_{\bm{n}_2 \sigma} h_{\bm{n}_3 \sigma} e_{\bm{n}_4 \sigma'} \Big]. \nonumber \\
\label{fermihubbardham}
\end{eqnarray}
where
\begin{equation}
t_i(\bm{n},\bm{n}') = \int d^2 r w_{i}^* ( \bm{r} - \bm{n}\lambda ) \left[  \frac{\hbar^2 }{2 m_i} \nabla^2
+ (-1)^{\delta_{ih}}  W_e (\bm{r}) \right] 
w_{i}( \bm{r} - \bm{n}'\lambda )
\label{tdefinitionfermi}
\end{equation}
\begin{align}
U_{ii}( \bm{n}_1,\bm{n}_2,\bm{n}_3,\bm{n}_4) & = \int d^2 r d^2 r' 
w_{i}^* ( \bm{r} - \bm{n}_1\lambda ) w_{j}^* ( \bm{r}' - \bm{n}_2\lambda )  V ( |\bm{r}-\bm{r}' | )  \nonumber \\
& \times w_{j} ( \bm{r}' - \bm{n}_3\lambda ) w_{i} ( \bm{r} - \bm{n}_4\lambda ) ,
\label{coulombdefinitionfermi} \\
U_{eh}( \bm{n}_1,\bm{n}_2,\bm{n}_3,\bm{n}_4) & = \int d^2 r d^2 r' 
w_{e}^* ( \bm{r} - (\bm{n}_1 + (1,1)/2 ) \lambda ) w_{h}^* ( \bm{r}' - \bm{n}_2\lambda )  V (\sqrt{|\bm{r}-\bm{r}'|^2 + d^2} )  \nonumber \\
& \times w_{h} ( \bm{r}' - \bm{n}_3\lambda ) w_{e} ( \bm{r} - ( \bm{n}_4+ (1,1)/2)\lambda ) ,
\end{align}
with $ i = e,h $ and $ \delta_{ih} $ is a Kronecker delta. The Wannier functions $ w_{i} ( \bm{r}) $ 
differ for electrons and holes due to their different masses. There is a lattice offset of $(1,1)/2 $ since we
assume a type II potential, i.e. the 
potential minima locations for electrons and holes differ by half a lattice unit. 
A minimal approximation to (\ref{fermihubbardham}) is 
to retain the nearest neighbor terms in (\ref{tdefinitionfermi}) and on-site terms
in (\ref{coulombdefinitionfermi}). Fig. \ref{fig:Uandtfermi} shows the results of our numerical evaluations of $ t_i = t_i((n_x,n_y),(n_x+1,n_y)) = t_i((n_x,n_y),(n_x,n_y+1)) $ and $ U_{ij} = U_{ij}( \bm{n},\bm{n},\bm{n},\bm{n}) $. 
In a similar way to Bose-Hubbard parameters of Fig. \ref{fig:Uandtplot}, the application of the periodic potential $ W_e (\bm{r}) $ acts to 
increase the electron-electron and hole-hole interaction and decrease the hopping amplitude. 
The electron-hole interaction plateaus off since potential minima of the two particle species sit on two spatially separate sublattices. Comparison with Ref. \cite{ogawa05} reveals that for $ U_{ij}/(t_e+t_h) \gg 1 $, 
and at a density of one exciton per site (half-filling in the terminology of Ref. \cite{ogawa05}), the excitons will be in a Mott-insulating regime in both the electron and hole layers. Thus for a large enough potential $ W_0^e $
the system will lie in such a Mott insulating phase. We again assume that the lifetimes of the indirect excitons (which can exceed $ \sim \mu$s according to Ref. \cite{voros05}) should exceed the requirements given in (\ref{superlifetime}) and (\ref{motttime}) in the respective phases.

\begin{figure}
\scalebox{0.6}{\includegraphics[bb=0 0 306 290]{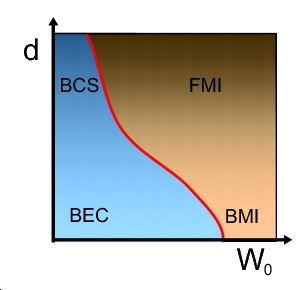}}
\caption{\label{fig:phasediagram} (Color online)  
Qualitative phase diagram of the electron-hole bilayer system at $n=1 $ unit filling density.  The labeled phases are BCS phase (BCS), Bose-Einstein condensate (BEC), Bose Mott insulator (BMI), and Fermi Mott insulator (FMI).  
Shading represents the transition from boson (light) to fermion (dark) behavior. Solid line denotes a first-order phase transition. }
\end{figure}

Examining various limits leads us to draw a qualitative phase diagram as shown in 
Fig. \ref{fig:phasediagram}.  First consider traveling up the $ d $ axis, with $ W_0 = 0 $. 
Assuming a periodic potential with $ \lambda = 0.1 $ $\mu$m, one exciton per potential minima corresponds
to a density of $ n_{\mbox{\tiny exc}} = 10^{10} $ $ \mbox{cm}^{-2} $. Monte Carlo calculations have 
predicted that Wigner crystallization should
occur at $ r_s \approx 37 $ \cite{tanatar89}, corresponding to a density of 
$ n_{\mbox{\tiny WC}} = 2.3 \times 10^8 $ $ \mbox{cm}^{-2} $ in GaAs.
As $ n_{\mbox{\tiny exc}} $ is above the Wigner crystal melting density $ n_{\mbox{\tiny WC}} $, we expect that the bilayer system should be conducting (i.e. a metallic phase) for $ d \rightarrow \infty $. 
For $ d=0 $, the system is still in a fairly 
low-density regime ($ a_B^2/\lambda^2 \ll 1 $), and thus we expect that the ground state may be 
described by a BEC (i.e. a non-localized metallic state) for sufficiently low temperatures. As $ d $ is increased, the Bohr radius of the excitons increase, until the exciton wavefunctions start to overlap. Beyond this point, the excitons cannot be 
described as bosons anymore, and the system enters a BCS phase \cite{comte82,littlewood96}. Moving in the direction of increasing $ W_0 $ for small $ d$, as discussed in section \ref{sec:motthubbard}, we expect a Bose-Hubbard transition into a Mott insulating phase. From the
considerations of Ref. \cite{ogawa05}, at unit exciton density we expect the system to be in a Mott 
insulating phase for $ U_{ee} > U_{eh} $, $ U_{hh} > U_{eh} $, and $ U_{ij}/(t_e+t_h) \gg 1 $. 
We thus expect that a transition should occur from the electron-hole plasma phase to a Mott insulating 
phase for large $ d $. Connecting the two boundaries for small and large $ d $ leads to the phase 
diagram Fig. \ref{fig:phasediagram}. It is plausible to expect that the Bose and Fermi Mott insulating states can be smoothly connected, in a similar way to a BEC-BCS crossover \cite{comte82,littlewood96}. We thus conjecture that the first order transition line between the metallic and Mott insulating states can also be smoothly connected throughout the phase
diagram. The repulsion between the particle species generally increases with increasing $ d $, 
as can be seen in Fig. \ref{fig:coulomb_d_dependence}.  Thus qualitatively the transition should shift to smaller values of $ W_0 $ for the fermionic 
limit, as shown in Fig. \ref{fig:phasediagram}.

\section{Summary and Conclusions}
\label{sec:conclusions}

We have considered the effect of applying a periodic potential on interacting exciton-polaritons and indirect excitons. 
Our main result is shown in Fig. \ref{fig:Uandtplot}, where the Bose-Hubbard parameters for the on-site interaction $ U$ and the tunneling amplitude $ t $ was calculated. 
We also derived a guideline (eq. (\ref{masscriterion})) for the range of parameters necessary to realize a phase transition from a BEC phase into a Mott insulating phase. The results
suggest that very exciton-like polaritons are required to observe the transition. 
Loosely speaking, the reason is that for the typical experimental parameters, the tunneling amplitude $ t $ is far greater than the interaction energy $ U $. Thus in order to make these parameters on the same order, the polariton mass needs to be increased to reduce $ t $. This results in the necessity of polaritons with a large exciton component. Alternatively, a very large potential amplitude $ W_0 $ can be applied. The experimental challenge in this case is to maintain $ U$ and $t$ greater than the experimental temperature and system disorder. Since the energy scale of the Hubbard parameters 
are set by the applied potential period,  this favors small $ \lambda $ in order to increase the energy scale. Although we focused mainly on parameters for GaAs, we note our formulas are general enough such that a simple substitution of material and geometrical parameters in Fig. \ref{fig:Uandtplot} should be enough to find the Hubbard parameters for any semiconductor system. 

We have also considered the effect of increasing the bilayer separation for indirect excitons, where there is a crossover
from a Bose-Hubbard model to a double Fermi-Hubbard model. The Hubbard parameters for the fermionic limit were 
derived (Fig. \ref{fig:Uandtfermi}). A Mott transition should be present for both limits, 
thus we argue that there should be a transition for all intermediate $ d $.  In an analogous way that there is a BEC-BCS crossover for zero potential \cite{comte82,littlewood96}, the Mott insulating limit should also crossover from a Bose Mott insulator to a double Fermi Mott insulator for large potentials. Our argument is based on connecting the various limits of the system, and requires a more rigorous numerical investigation to confirm our conjecture.  A more detailed investigation of the various phases would
require an extensive numeric survey of the parameter space, which we leave as future work.

\section{Acknowledgements}

This work is supported by the Special Coordination Funds for Promoting Science and Technology, Navy/SPAWAR Grant N66001-09-1-2024, and MEXT.  P.R. would like to acknowledge financial support from the German Research Foundation (DFG) via Tr950/1-1 and Re2978/1-1. T. B. and P. R. thank Hui Deng, David Press, and Sven H{\"o}fling for valuable comments regarding the manuscript.

\appendix

\section{Effective polariton interaction}

\subsection{Exciton-exciton interaction}
\label{app:effectiveinteraction}

Eq. (\ref{directtermdefinition}) may be evaluated by making a change of variables to $ \bm{R}= \beta_e \bm{r}_e + \beta_h \bm{r}_h $ and 
$ \bm{\rho}= \bm{r}_e -  \bm{r}_h $, after which we obtain
%
\begin{align}
\label{directtermone}
U_{\mbox{\tiny dir}} (\bm{Q}, \bm{Q}',\bm{q}) & = 
\left[ -2E_{1s} + \frac{\hbar^2}{2M} (\bm{Q}^2 + \bm{Q}'^2) 
\right]\delta( \bm{q}) \nonumber \\
& + \frac{N_G^4 e^2}{4 \pi \epsilon A a_B^4} \exp \left( \frac{2d \lambda(d)}{a_B} \right) 
\int d^2 \rho d^2 \rho'  \frac{2 \pi}{q} \big[  e^{-i \bm{q} \cdot \beta_h ( \bm{\rho} - \bm{\rho}' ) } + 
e^{i \bm{q} \cdot \beta_e ( \bm{\rho} - \bm{\rho}' ) } \nonumber \\
& - e^{-dq - i \bm{q} \cdot ( \beta_h \bm{\rho} +  \beta_e \bm{\rho}' ) } - e^{-dq + i \bm{q} \cdot ( \beta_e \bm{\rho} +  \beta_h \bm{\rho}' ) }
\big] e^{-  \lambda ( d) (\sqrt{\rho^2 + d^2} + \sqrt{{\rho'}^2 + d^2} )/a_B } .
\end{align} 
Eq. (\ref{finaldirect}) may be obtained by performing the $ \rho $ and $ \rho' $ integrals separately and using the rotational invariance of $ \bm{q} $. 
Figure \ref{fig:direct} is obtained by numerically evaluating
\begin{equation}
I_{\mbox{\tiny dir}} (q,d) = \frac{2 \pi^5}{q a_B} N_G^4 \exp( 2 \frac{d}{a_B} \lambda(d) )  \left[  I_0(q,\beta_h,d)^2 + I_0(q,\beta_e,d)^2
-2 e^{-dq} I_0(q,\beta_e,d) I_0(q,\beta_h,d) \right],
\end{equation}
where
\begin{eqnarray}
I_0 (q,\beta,d) =  \int dr r J_0 ( q a_B \beta r) \exp \left[ -\lambda(d) \sqrt{r^2 + (d/a_B)^2} \right],
\end{eqnarray}
and $ J_0 ( x) $ is the Bessel function of the first kind.  The normalization assuming the electrons and holes are confined 
as delta-functions in the $z$-direction is
%
\begin{equation}
N_G = \sqrt{\frac{\lambda(d)^2}{2 \pi (1+  d \lambda(d)/a_B )}} .
\end{equation}

The electron and hole exchange terms may be obtained by following the derivation given in the Appendix B of Ref. 
\cite{ciuti98}. We obtain (\ref{electronexchange}) and (\ref{holeexchange}), where
\begin{eqnarray}
I_{\mbox{\tiny exch}} ( \Delta Q,q,\theta,\beta,d) & = & 
\left( \frac{\pi}{2} \right)^2  N_G^4 \exp \left( \frac{2 \lambda(d)d}{a_B} \right) \nonumber \\
& & \times \int_0^\infty dx \int_0^{2\pi} d \theta_x
\int_0^\infty dy_1 \int_0^{2\pi} d \theta_1 \int_0^\infty dy_2 \int_0^{2\pi} d \theta_2 
x y_1 y_2 \nonumber \\
& & \times \cos \{ \Delta Q a_B [ \beta x \cos (\theta-\theta_x) + \beta y_1 \cos (\theta-\theta_1) ]  \nonumber \\
& & + q a_B [ -x \cos \theta_x -\beta y_1  \cos \theta_1 + (1-\beta) y_2
\cos \theta_2] \}  \nonumber \\
& & \times \exp \Big( - \frac{\lambda(d)}{2} [ (y_2 \cos \theta_2 - y_1 \cos \theta_1 - x \cos \theta_x)^2 \nonumber \\
& & + (y_2 \sin \theta_2 - y_1 \sin \theta_1 - x \sin \theta_x)^2 + (d/a_B)^2 ]^{1/2} \Big) \nonumber \\
& & \exp \left( - \frac{\lambda(d)}{2} \sqrt{x^2 + (d/a_B)^2} \right) 
\exp \left( - \frac{\lambda(d)}{2} \sqrt{y_1^2 + (d/a_B)^2} \right) 
 \nonumber \\
& & \exp \left( - \frac{\lambda(d)}{2} \sqrt{y_2^2 + (d/a_B)^2} \right) \Big\{ \frac{1}{\sqrt{y_1^2 + x^2 + 2 y_1 x \cos ( \theta_1 - \theta_x)}} 
 \nonumber \\
& & + \frac{1}{\sqrt{y_2^2 + x^2 - 2 y_2 x \cos ( \theta_2 - \theta_x)}} -\frac{1}{\sqrt{y_2^2 + (d/a_B)^2}} -\frac{1}{\sqrt{y_2^2 + (d/a_B)^2}}
\Big\} \nonumber \\
\label{iexchexpression}
\end{eqnarray}

\subsection{Saturation interaction}
\label{app:saturationinteraction}

After substituting the exciton wavefunctions (\ref{excitonwavefunc}) into the expression  for the saturation interaction (\ref{saturationintegral}), we obtain
\begin{align}
U_{\mbox{\tiny sat}} (\bm{Q},\bm{Q}',\bm{q}) & = - \frac{\cal G}{A^{3/2}} \left( \sqrt{\frac{2}{\pi a_B^2}} \right)^3 
\int d^2 x d^2 r_e d^2 r_h 
\exp[-\frac{1}{a_B} \left( | \bm{r}_e - \bm{x} | +  | \bm{r}_h - \bm{x} |
+| \bm{r}_e - \bm{r}_h | \right)] \nonumber \\
& \exp[ i \bm{Q} \cdot (\beta_e \bm{r}_e + \beta_h \bm{r}_h) + i \bm{Q}' 
\cdot (\beta_e \bm{x} + \beta_h \bm{r}_h) - (\bm{Q} + \bm{Q}' - \bm{q})\cdot ( \beta_e \bm{r}_e + \beta_h \bm{r}_h ) - \bm{q} \cdot \bm{x} ] .
\end{align}
Changing variables to $ \bm{\xi} = (\bm{r}_e - \bm{r}_h)/(2 a_B) $, $ \bm{\eta} = ( (\bm{r}_e + \bm{r}_h)/2 - \bm{x} )/a_B $, and $ \gamma = \bm{x}/a_B $, 
\begin{align}
U_{\mbox{\tiny sat}} (\bm{Q},\bm{Q}',\bm{q}) & = - {\cal G} \frac{ a_B}{\sqrt{A} }  \frac{8 \sqrt{2}}{\pi^{3/2}} 
\int d^2 \xi d^2 \eta 
\exp [ - | \xi + \eta | - | \eta - \xi | - 2 |\xi | ] \nonumber \\
& \exp[ i a_B \bm{\xi} \cdot ( \beta_h \bm{Q} - \beta_e \bm{Q}' + (\beta_e - \beta_h) \bm{q} ) 
+ i a_B \bm{\eta} \cdot ( \bm{q} - \beta_h \bm{Q} - \beta_e \bm{Q}') ].
\end{align}
The dimensionless integral appearing in (\ref{saturationintegral2}) is 
\begin{align}
I_{\mbox{\tiny sat}} (\bm{Q},\bm{Q}',\bm{q})  & =  \frac{8 \sqrt{2}}{\pi^{3/2}} 
\int d \xi d \theta_{\xi} d \eta d \theta_{\eta} \xi \eta \nonumber \\
& \times \cos \{ a_B \xi [ \beta_h Q \cos ( \theta - \theta_{\xi} ) 
- \beta_e Q' \cos ( \theta' - \theta_{\xi} ) + (\beta_e - \beta_h)  q \cos \theta_{\xi} ] \nonumber \\
& + a_B \eta \left[ - \beta_h Q \cos ( \theta - \theta_{\eta} ) - \beta_e Q' \cos ( \theta' - \theta_{\eta}) + q \cos \theta_{\eta} \right] \} \nonumber \\
& \times \exp \left[ - \sqrt{\xi^2 + \eta^2 + 2\xi \eta \cos ( \theta_{\xi} - \theta_{\eta} ) }
-\sqrt{ \xi^2 + \eta^2 - 2 \xi \eta \cos ( \theta_{\xi} - \theta_{\eta} ) } - 2 \xi \right],
\end{align}
where $ \theta $ ($\theta$') is the angle between $\bm{q} $ and $ \bm{Q} $ ($\bm{Q}'$).

\section{Normalization of exciton number states}
\label{app:bosefermi}

By definition, we have
\begin{equation}
N(n) = \frac{1}{n!} \langle 0 | (b)^n (b^\dagger)^n | 0 \rangle .
\end{equation}
Using the commutation relation (\ref{bcommutation}), we obtain
\begin{equation}
\label{nofnexpression}
N(n) = N(n-1) - \frac{1}{n!} \sum_{m=0}^{n-1} \langle 0 | (b)^{n-1} (b^\dagger)^m
K (b^\dagger)^{n-1-m} | 0 \rangle.
\end{equation}
To obtain a simplified expression for the second term, it is useful to define
\begin{equation}
\label{kcommutation}
[K,b^\dagger] = J^\dagger
\end{equation}
where 
\begin{eqnarray}
J^\dagger &= &2 \int d^2 r_e d^2 r_e' d^2 r_h d^2 r_h'
\phi_d ( \bm{r}_e - \bm{r}_h' ) \phi_{d} ( \bm{r}_e' - \bm{r}_h' ) \phi_{d} ( \bm{r}_e' - \bm{r}_h ) \nonumber \\
& & \times w  ( \beta_e \bm{r}_e + \beta_h \bm{r}_h' ) 
w  ( \beta_e \bm{r}_e' + \beta_h \bm{r}_h' ) 
w  ( \beta_e \bm{r}_e' + \beta_h \bm{r}_h ) e^\dagger (\bm{r}_e) h^\dagger (\bm{r}_h) .
\end{eqnarray}
Retaining only powers linear in the operator $ K$, we obtain
\begin{equation}
\label{kidentity}
\langle 0 | (b^\dagger)^n (b)^m K (b)^{n-m} | 0 \rangle = (n-m) n! \langle 0 | b K b^\dagger | 0 \rangle 
\end{equation}
Substituting this into (\ref{nofnexpression}), we obtain
\begin{equation}
N(n) = N(n-1) - \frac{1}{2}(n-1) \langle  0 | b K b^\dagger | 0 \rangle .
\end{equation}
This definition may be used recursively, to obtain the final result (\ref{normalizationfactor}). 

The matrix element above may be calculated according to
\begin{eqnarray}
\langle 0 | b K b^\dagger | 0 \rangle & \equiv & I_{\mbox{\tiny BF}} (W_0,d) = 2 \int 
d^2 r_e d^2 r_h d^2 r_e' d^2 r_h'  \phi_{d} ( \bm{r}_e - \bm{r}_h )
 \phi_{d}  ( \bm{r}_e - \bm{r}_h' )  \phi_{d}  ( \bm{r}_e' - \bm{r}_h )
 \phi_{d}  ( \bm{r}_e' - \bm{r}_h' ) \nonumber \\
& & \times w ( \beta_e \bm{r}_e + \beta_h \bm{r}_h )
w ( \beta_e \bm{r}_e + \beta_h \bm{r}_h' )
w ( \beta_e \bm{r}_e' + \beta_h \bm{r}_h )
w ( \beta_e \bm{r}_e' + \beta_h \bm{r}_h' ) .
\end{eqnarray}
Making the transformation to center of mass coordinates $ \bm{z}_r = \bm{r}_e - \bm{r}_h  $, $ \bm{z}_r' = \bm{r}_e - \bm{r}_h' $, $ \bm{z}_r = \bm{r}_e' - \bm{r}_h $, $ \bm{z}_{\mbox{\tiny CM}} = \beta_e (\bm{r}_e + \bm{r}_e')/2 + \beta_h (\bm{r}_h + \bm{r}_h')/2 $, we obtain
\begin{equation}
\label{idefinition}
I_{\mbox{\tiny BF}} (W_0,d) \approx 2 \int d^2 z_{\mbox{\tiny CM}} | w ( \bm{z}_{\mbox{\tiny CM}} ) |^4 
\int d^2 z_r d^2 z_r' d^2 z_r''  \phi_{d}  (\bm{z}_r)  \phi_{d}  (\bm{z}_r')  \phi_{d}  (\bm{z}_r'') 
 \phi_{d}  (\bm{z}_r''-\bm{z}_r +\bm{z}_r') ,
\end{equation}
where we have used the fact that the relative wavefunction $ \phi ( \bm{r}) $ extends out
to a distance of the order of $ \sim a_B $, 
while the Wannier function extends out to at a distance $ \sim \lambda$, with $ \lambda \gg a_B $. The Wannier function has dimensions of 
the inverse length (in 2D), hence the order of magnitude of  the first integral is  $ \sim 1/\lambda^2 $. The order of magnitude of  the second integral is $ \sim a_B^2 $, making the whole integral of the order of $ \sim  a_B^2/\lambda^2 $. As can be shown by direct calculation, integrals 
involving higher powers in the operator $ K $ involve higher powers of $  a_B^2/\lambda^2 $. Therefore, 
the approximation made in (\ref{kidentity}) is thus reasonable as long as $ a_B^2 \ll \lambda^2 $.


\clearpage

\begin{thebibliography}{99}

\bibitem{deng02}
H. Deng {\it et al.}, Science {\bf 298}, 199 (2002).

\bibitem{kasprzak06}
J. Kasprzak {\it et al.}, Nature {\bf 443}, 409 (2006). 

\bibitem{balili07}
R. Balili {\it et al.}, Science {\bf 316}, 1007 (2007). 

\bibitem{deng06}
H. Deng {\it et al.}, Phys. Rev. Lett. {\bf 97}, 146402 (2006). 

\bibitem{amo09}
A. Amo {\it et al.}, Nature {\bf 457}, 291 (2009).

\bibitem{lagoudakis08} 
K. Lagoudakis {\it et al.}, Nat. Phys. {\bf 4}, 706 (2008). 

\bibitem{roumpos09}
G. Roumpos {\it et al.}, submitted.  

\bibitem{utsunomiya08}
S. Utsunomiya {\it et al.}, Nat. Phys. {\bf 4}, 700 (2008).

\bibitem{greiner02}
M. Greiner {\it et al.}, Nature {\bf 415}, 39 (2002). 

\bibitem{feynman82}
R. Feynman, Int. J. Theo. Phys. {\bf 21}, 467 (1982).

\bibitem{buluta09}
I. Buluta and F. Nori, Science {\bf 326}, 108 (2009).

\bibitem{hartmann06}
M. J. Hartmann, F. G. S. L. Brand{\~a}o, and M. B. Plenio, Nat. Phys. {\bf 2}, 849 (2006).

\bibitem{greentree06}
A. D. Greentree, C. Tahan, J. H. Cole, and L. C. L. Hollenberg,  Nat. Phys. {\bf 2}, 856 (2006). 

\bibitem{lai07}
C. W. Lai {\it et al.},  Nature {\bf 450}, 529 (2007). 

\bibitem{kim08}
N. Y. Kim {\it et al.}, Phys. Stat. Sol. {\bf 245}, 1076 (2008). 

\bibitem{daif06}
O. El Daif {\it et al.}, Appl. Phys. Lett. {\bf 88}, 061105 (2006).

\bibitem{na08}
N. Na, S. Utsunomiya, L. Tian, and Y. Yamamoto, Phys. Rev. A {\bf 77}, 031803(R) (2008). 

\bibitem{na09}
N. Na and Y. Yamamoto, arxiv: 0804.1829.

\bibitem{aichorn08}
M. Aichhorn, M. Hohenadler, C. Tahan, and P. B. Littlewood, Phys. Rev. Lett. {\bf 100}, 216401 (2008).

\bibitem{koch09}
J. Koch and K. Le Hur, Phys. Rev. A {\bf 80}, 023811 (2009). 

\bibitem{grochol09}
M. Grochol, Phys. Rev. B {\bf 79}, 205306 (2009).

\bibitem{keldysh95}
L. V. Keldysh, in {\it Bose-Einstein Condensation}, edited by A. Griffin, D. W. Snoke, and S. Stringari
(Cambridge University Press, Cambridge, UK, 1995) and references therein. 

\bibitem{butov02}
L. V. Butov, C. W. Lai, A. L. Ivanov, A. C. Gossard, and D. S. Chemla, Nature {\bf 417}, 47 (2002).

\bibitem{remeika09}
M. Remeika {\it et al.}, Phys. Rev. Lett. {\bf 102}, 186803 (2009).

\bibitem{rapaport05}
R. Rapaport {\it et al.}, Phys. Rev. B {\bf 72}, 075428 (2005).

\bibitem{hammack06}
A. T. Hammack {\it et al.}, Appl. Phys. Lett. {\bf 99}, 066104 (2006).

\bibitem{rudolph07}
J. Rudolph, R. Hey, and P. V. Santos, Phys. Rev. Lett. {\bf 99}, 047602 (2007). 

\bibitem{lima05}
M. M. de Lima Jr. and P. V. Santos, Rep. Prog. Phys. {\bf 68}, 1639 (2005). 

\bibitem{byrnes07}
T. Byrnes, P. Recher, N. Y. Kim, S. Utsunomiya, and Y. Yamamoto, Phys. Rev. Lett. {\bf 99}, 016405 (2007).

\bibitem{deleon01}
S. B. de-Leon and B. Laikhtman, Phys. Rev. B {\bf 63}, 125306 (2001). 

\bibitem{hanamura79}
E. Hanamura and H. Haug, Phys. Rep. {\bf 33C}, 209 (1979). 

\bibitem{rochat00}
G. Rochat, C. Ciuti, V. Savona, C. Piermarocchi, A. Quattropani, and P. Schwendimann, Phys. Rev. B {\bf 61}, 13856 (2000). 

\bibitem{stolz81}
H. Stolz, R. Zimmerman, and G. R{\"o}pke, Phys. Status Solidi B {\bf 105}, 585 (1981). 

\bibitem{bobrysheva72}
A. I. Bobrysheva, M. F. Miglei, and M. I. Shmiglyuk, Phys. Status Solidi B {\bf 53}, 71 (1972). 

\bibitem{ciuti98}
C. Ciuti, V. Savona, C. Piermarocchi, A. Quattropani, and P. Schwendimann, Phys. Rev. B {\bf 58}, 7926 (1998). 

\bibitem{deng07}
H. Deng, G. S. Solomon, R. Hey, K. H. Ploog, and Y. Yamamoto, Phys. Rev. Lett. {\bf 99}, 126403 (2007). 

\bibitem{leavitt90}
R. P. Leavitt and J. W. Little, Phys. Rev. B {\bf 42}, 11774 (1990). 

\bibitem{sheshadri93}
K. Sheshadri, H. R. Krishnamurthy, R. Pandit, and T. V. Ramakrishnan, Europhys. Lett. {\bf 22}, 257 (1993).

\bibitem{comte82}
C. Comte and P. Nozi{\`e}res, J. Physique {\bf 43}, 1069 (1982). 

\bibitem{littlewood96}
P. B. Littlewood and X. Zhu, Phys. Scripta T{\bf68}, 56 (1996). 

\bibitem{ogawa05}
T. Ogawa, Y. Tomio, and K. Asano, J. Phys. Conf. Series {\bf 21}, 112 (2005). 

\bibitem{tanatar89}
B. Tanatar and D. M. Ceperley, Phys. Rev. B {\bf 39}, 5005 (1989).

\bibitem{tomio06}
Y. Tomio, K. Honda, and T. Ogawa, Phys. Rev. B {\bf 73}, 235108 (2006).

\bibitem{feldmann87}
J. Feldmann {\it et al.}, Phys. Rev. Lett. {\bf 59}, 2337 (1987). 

\bibitem{voros05}
Z. V{\"o}r{\"o}s, R. Balili, D. W. Snoke, L. Pfeiffer, and K. West, Phys. Rev. Lett. {\bf 94}, 226401 (2005). 



\end{thebibliography}
\end{document}